\documentclass[12pt, amsmath, amssymb]{iopart}

\usepackage{mathrsfs}
\usepackage{graphicx}
\usepackage{subfigure}
\usepackage{amssymb}
\usepackage{epsf,latexsym}
\usepackage{times}
\usepackage{ifpdf}
\ifpdf
\usepackage{epstopdf}   
\fi

\usepackage{color}

\newcommand{\bra}[1]{\langle#1|}
\newcommand{\ket}[1]{|#1\rangle}

\newcommand{\be}{\begin{equation}}
\newcommand{\ee}{\end{equation}}
\newcommand{\bea}{\begin{eqnarray}}
\newcommand{\eea}{\end{eqnarray}}

\usepackage{color}

\begin{document}

\bibliographystyle{unsrt}

\title{Measurement-based quantum computing with a spin ensemble coupled to a stripline cavity}

\author{Yuting Ping}
\ead{yuting.ping@materials.ox.ac.uk}
\address{Department of Materials, University of Oxford, Parks Road, Oxford, OX1 3PH, UK}

\author{Erik M. Gauger}
\address{Centre for Quantum Technologies, National University of Singapore, 3 Science Drive 2, Singapore 117543}
\address{Department of Materials, University of Oxford, Parks Road, Oxford, OX1 3PH, UK}

\author{Simon C. Benjamin}
\ead{simon.benjamin@materials.ox.ac.uk}
\address{Department of Materials, University of Oxford, Parks Road, Oxford, OX1 3PH, UK}
\address{Centre for Quantum Technologies, National University of Singapore, 3 Science Drive 2, Singapore 117543}



\begin{abstract}
Recently a new form of quantum memory has been proposed. The storage medium is an ensemble of electron spins, coupled to a stripline cavity and an ancillary readout system. Theoretical studies suggest that the system should be capable of storing numerous qubits within the ensemble, and an experimental proof-of-concept has already been performed. Here we show that this minimal architecture is not limited to storage but  is in fact capable of full quantum processing by employing measurement-based entanglement. The technique appears to be remarkably robust against the anticipated dominant error types. The key enabling component, namely a readout technology that non-destructively determines ``are there $n$ photons in the cavity?'', has already been realised experimentally.
\end{abstract}

\maketitle

\section{Introduction}
\begin{figure}[b]
\begin{center}$
\begin{array}{cc}
  \subfigure[Circuit diagram of the integrated device for ensemble QIP.]{\label{fig:meas} \includegraphics[width=3.6in]{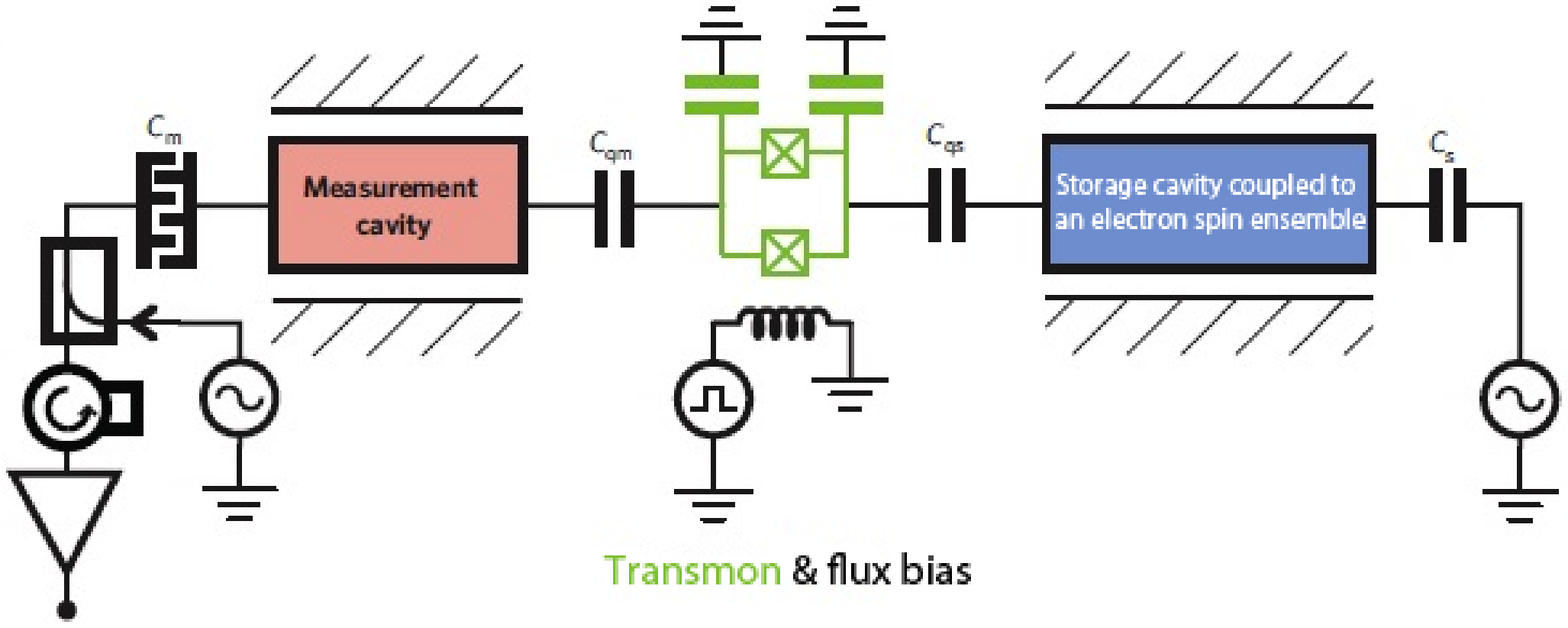}} & \hspace{3mm}
  \subfigure[Storage cavity coupled with an electron spin ensemble.]{\label{fig:storage} \includegraphics[width=2in]{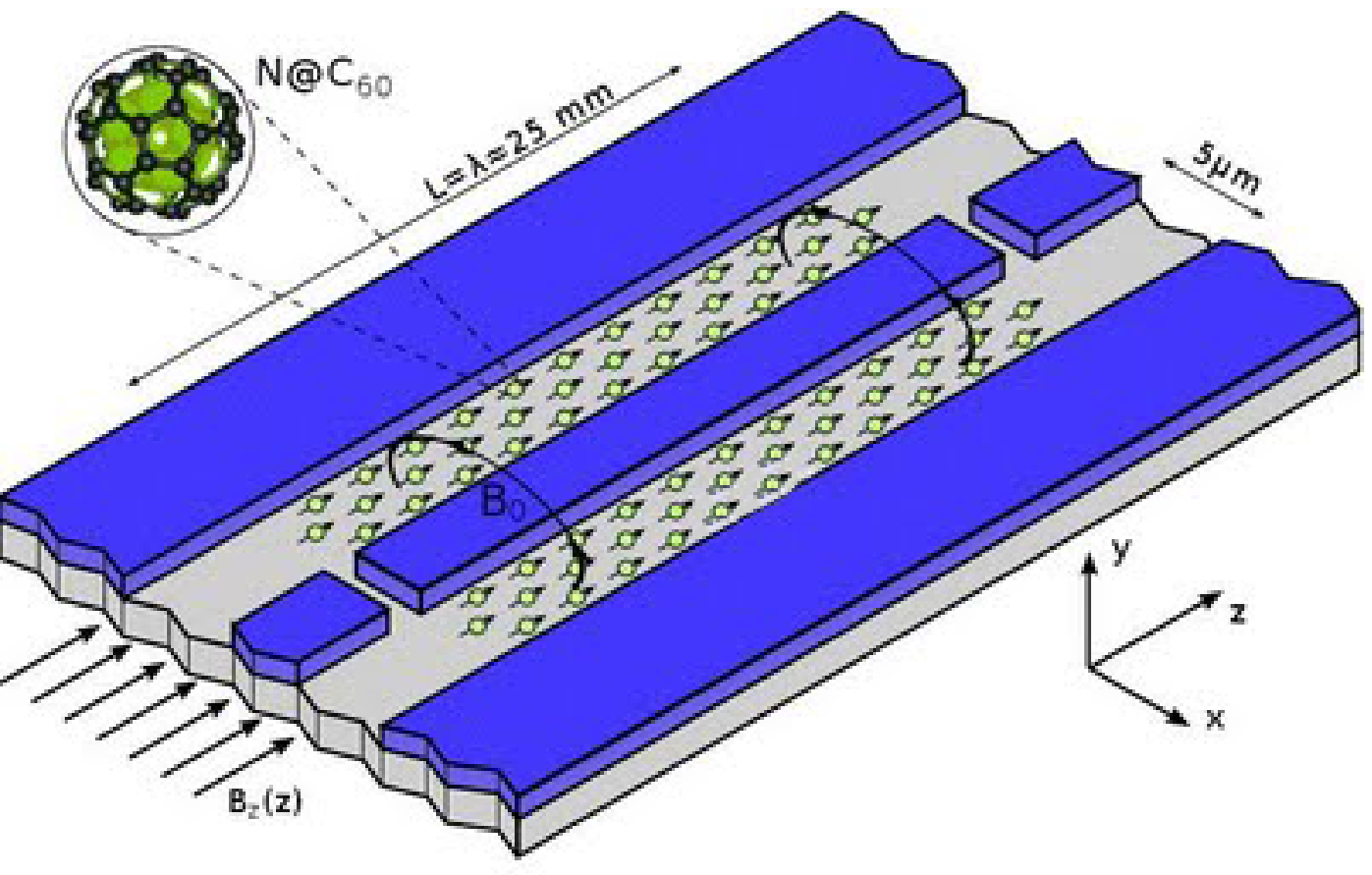}}
\end{array}$
\caption{(a) Circuit diagram of the integrated device for ensemble QIP, with figure adapted from ~\cite{johnson10} (see details of physical implementation on a chip in~\cite{johnson10}). The number of photons in storage cavity affects the transmon state, which is indicated from microwave probing the measurement cavity; (b) Physical setup of the storage line cavity coupled with an ensemble of $N= 10^{11}$ electron spins (N@C$_{60}$) doped on substrate, with an average coupling $\bar{g} \simeq 2 \pi \times 20$ Hz. A bias field of 180 mT is required to bring the spin Larmor precession in resonance with the cavity at a resonance frequency of $2 \pi \times 5$ GHz~\cite{wesenberg09}. A switchable linear magnetic gradient field for appropriate time lengths (gradient pulses) is required in order to access the different collective modes of the ensemble. Figure adapted from~\cite{wesenberg09}, with the CPB qubit removed.}
\label{fig:device}
\end{center}
\end{figure}

Interactions concerning single spins are usually very weak, which imposes a huge challenge on the feasibility of their manipulation and measurement, and hence their potential for quantum information processing ({\it QIP}). Inspired by earlier work~\cite{rabl06, andre06}, it was suggested in 2009 that electron spin ensembles on the other hand could exhibit strong couplings with a collective field via magnetic dipoles \big(see Fig.~\ref{fig:storage}\big)~\cite{wesenberg09}, exploiting the extremely small mode volumes in stripline cavities. Such strong couplings have also been recently demonstrated by independent experimental groups~\cite{kubo10, schuster10a}. 

Moreover, it was further proposed that the various spin waves in the ensemble could be used as independent quantum registers~\cite{wesenberg09}, for which a proof-of-principle experiment was carried out demonstrating the storage and readout of multiple classical excitations~\cite{wu10}. Therefore, so-called holographic QIP schemes~\cite{tordrup08a, nunn08, afzelius10} should be possible by selectively coupling modes of the spin ensemble to the superconducting transmission line cavity~\cite{wesenberg09}. However, promoting a memory system to a full QIP device is non-trivial because the allowed operations in such a system are analogous to those in linear optics, and thus the provision of entangling gates requires non-linearity supplied by other components. One suggestion involves integrating a transmon Cooper pair box ({\it CPB}) \cite{koch07} in the cavity; moving a memory qubit to the CPB would facilitate universal qubit operations, as well as the readout procedures~\cite{wesenberg09, tordrup08a, wallraff05}.

However, CPBs are not ideal for representing qubits due to their short coherence times~\cite{koch07}. Therefore in this paper we propose a composite system (see Fig.~\ref{fig:device}) where such devices are only used for measurement, but in such a fashion that in addition to simple readout, also entangling projections can be implemented. We envisage the use of a system that is both number resolving~\cite{schuster07} and also non-destructive;  recent experimental demonstrations of such systems have been accomplished~\cite{johnson10, reed10}. The realisation consists of two cavities coupled via a transmon CPB; measurement of the number of single photons in the storage cavity is achieved via a read-out procedure involving the second cavity~\cite{johnson10}. In effect this allows one to ask the question, {\it are there exactly $n$ photons in the storage cavity}? If we wish to subsequently ask about a different $n$, we would employ a flux bias to tune the transmon frequency (a nanosecond timescale process). In the ideal case when no errors are present, if the result is NO then any coherent superposition of photon number states other than $\ket{n}$ is preserved in the storage cavity~\cite{johnson10}, which is coupled to the electron spin ensemble~\cite{wesenberg09}. Appropriate bias field and magnetic gradient pulses are applied for resonantly accessing particular modes of the ensemble as discussed in Ref.~\cite{wesenberg09}. 

In contrast to previous proposals, here the CPB no longer performs the role of gate operations on the mode qubits nor to store the qubits. After a brief review of the basic physics in the spin ensemble we shall show that, by using {\it dual-rail} encoding, a universal set of quantum gates can be implemented for the logical qubits. Any single qubit rotation can be achieved by applying appropriate magnetic gradient pulses and adjusting the bias field when necessary. Importantly, the two-qubit parity projection that we shall presently discuss can enable general quantum computing through, for example, the creation of graph states~\cite{raussendorf01}.

\section{Modelling Collective Mode-Cavity Coupling}
Suppose the ensemble of $N$ electron spins is in its ground state ${\bf \ket{0}} = \ket{0...0}$ and that the cavity contains a single microwave photon. If the bias field is such that the spins' Zeeman splitting is resonant with the microwave photon, then the ensemble will collectively absorb that photon on a timescale that is proportional to $1/\sqrt{N}$. The ensemble state after the photon absorption is then given by
\begin{equation}
\ket{\psi_1 (0)} = \frac{1}{\sqrt{N}} \sum_q \frac{g_q}{\bar{g}} \ket{0_1...1_q...0_N}
\label{eq:cite1}
\end{equation}
where the sum is over all possible spin-flip (``1") positions $q$ in the ensemble (see~\cite{wesenberg09}). Here, $g_q$ is the cavity coupling strength with the $q^{th}$ electron spin in the ensemble, and $\bar{g}$ is the average strength. The above state and state ${\bf \ket{0}}$ together form an effectively closed two-level system (i.e., a mode qubit)~\cite{wesenberg09}. Now if no parameters are changed then the quantum of energy will ``flip-flop" back and forth between the cavity and the collective state. 

However, if we wish to stop the flip-flopping, i.e., decouple the cavity from this memory mode, then we will temporarily apply a linear gradient in the magnetic field such that each spin will acquire phase at a different rate. If the gradient pulse causes the field to vary in the $z$ direction, and writing the coordinate of the $q^{th}$ spin as $z_q$, then the collective state becomes~\cite{wesenberg09} 
\begin{equation}
\ket{\psi_1 (k)} = \frac{1}{\sqrt{N}} \sum_q \frac{g_q}{\bar{g}} e^{i k (\xi) z_q} \ket{0_1...1_q...0_N}.
\label{eq:cite2}
\end{equation}
Here, parameter $k$ depends on the strength of the gradient and its duration $\xi$, and consequently the various terms in Eq.~\ref{eq:cite2} have developed relative phases with one another~\cite{wesenberg09}. This prevents the $\sqrt{N}$-enhanced mode-cavity coupling, and a single excitation is thus stored in the spin ensemble. Indeed, if we were to introduce a new photon into the cavity at this stage then that photon will resonantly transfer to the ensemble almost independently of the presence of the former excitation~\cite{wesenberg09}. After the application of the second gradient pulse, this procedure can be repeated yet again.

As long as the number of the excitations $n \ll N$ with an upper limit of $n_{max} \leq \sqrt{N}$ (since the coupling enhancement of $\sqrt{N-n} \sim \sqrt{N}$ is still very large for small $n$), single excitations can be {\it independently} stored into and read out of the different collective modes $i$ by appropriately applying $\pm k_i$-pulses 
with $k_i = k (\xi_i)$~\cite{wesenberg09}. The superradiant state $\ket{\psi_1 (0)} $ corresponds to the $k = 0$ mode, and each time a $k_i$-pulse is applied it maps the $k = 0$ mode to the $i^{th}$ mode. Therefore, to access a particular mode $i$ a $- k_i$-pulse is applied to map it to the $k = 0$ mode, which then interacts with the cavity strongly upon resonance. A $k_i$-pulse is then applied to map the mode back. To stop the the mode-cavity coupling, one can simply tune the bias field such that the cavity is out of resonance with the ensemble spins~\cite{wesenberg09}.

Therefore, the Hamiltonian coupling a particular mode $i$ with the cavity for a single excitation with energy $\epsilon$ is 
\begin{equation}
H_i^{(1)} = \epsilon\ (m_i^{\dagger} m_i + c^{\dagger} c) + J (m_i^{\dagger} c + m_i c^{\dagger}) =
\left( {\begin{array}{cc}
\epsilon & J \\
J & \epsilon  \\
\end{array}}\right)
\end{equation}
in the basis of $\ket{1_M 0_C}$, $\ket{0_M 1_C}$. Here, $m_i^{\dagger}$ ($c^{\dagger}$) and $m_i$ ($c$) are the corresponding mode excitation (cavity photon) creation and annihilation operators, and $J = \sqrt{N} \bar{g}$ is the effective coupling strength being $\sim 2 \pi \times 6$ MHz for the parameters in Fig.~\ref{fig:storage}. This effective coupling $J$ is orders of magnitude larger than the decay rates of collective spin excitations and the cavity decay rate~\cite{wesenberg09}. The corresponding time evolution operator for a single excitation is then
\begin{equation}
S^{(1)} (t) = e^{-i \epsilon t}
\left( {\begin{array}{cc}
\cos Jt & - i \sin Jt \\
- i \sin Jt & \cos Jt  \\
\end{array}}\right).
\label{eq:unitary}
\end{equation}
After a time $t= \tau : = \frac{\pi}{2 J} \simeq 40$ ns, a full {\it SWAP} of one single excitation has occurred between the cavity and mode $i$, whereas for $t \leq \tau$ only a partial {\it SWAP} operation has taken place. 

When two or more excitations $n$ exist between a particular mode $i$ interacting with the cavity, the coupling strength and thus the time required for an exchange are adjusted by a factor of $\sqrt{n}$. The cavity-mode Hamiltonian for two excitations is then
\begin{equation}
H_i^{(2)} =
\left( {\begin{array}{ccc}
2 \epsilon & \sqrt{2} J & 0 \\
\sqrt{2} J  & 2 \epsilon & \sqrt{2} J   \\
0 & \sqrt{2} J  & 2 \epsilon \\
\end{array}}\right)
\label{eq:2ex}
\end{equation}
in the basis of $\ket{2_{M} 0_C}$, $\ket{1_{M} 1_C}$, $\ket{0_{M} 2_C}$, with the corresponding time evolution operator
\begin{equation}
S^{(2)} (t) = e^{-2 i \epsilon t}
\left( {\begin{array}{ccc}
\cos^2 Jt & - \frac{i}{\sqrt{2}} \sin 2Jt & - \sin^2 Jt \\
- \frac{i}{\sqrt{2}} \sin 2Jt  & \cos 2Jt & - \frac{i}{\sqrt{2}} \sin 2Jt  \\
- \sin^2 Jt & - \frac{i}{\sqrt{2}} \sin 2Jt  & \cos^2 Jt \\
\end{array}}\right).
\label{eq:unitary2}
\end{equation}
Note that a full {\it SWAP} of two excitations still occurs after the same time $t = \tau$ as for the single excitation case. 

For simplicity, from this point on whenever we say a particular mode $i$ is interacting with the cavity for some time $t$, a $- (+) k_i$-pulse is by default applied right before (after) the interaction which has a duration of $t$ with the bias field necessary for resonance. Note that the time for the necessary gradient pulses can be shortened by increasing the gradient strength~\cite{wesenberg09}.

\section{Single Qubit Operations in Dual-Rail Encoding} 
\begin{figure}[h]
\begin{center}$
\begin{array}{cc}
  \subfigure[X-type rotations: {\it X} gate when $t = \tau$]{\label{fig:x} \includegraphics[width=2.5in]{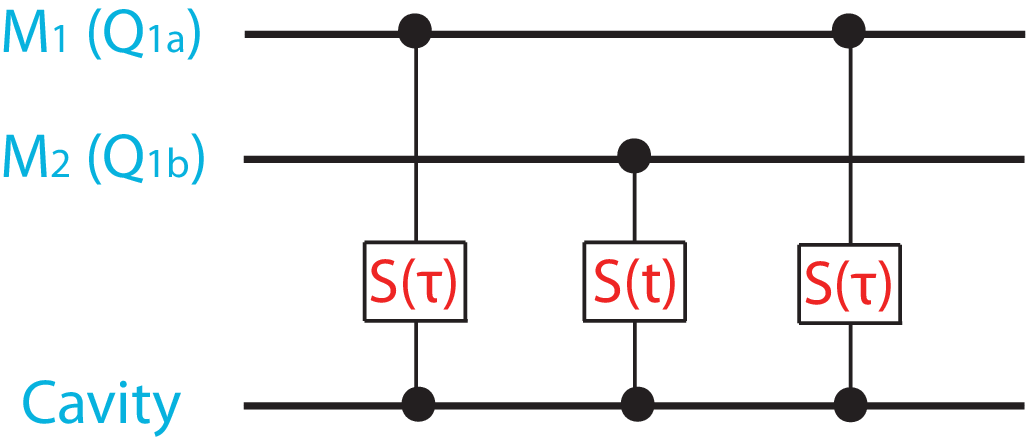}} & \hspace{5mm}
  \subfigure[Z rotations: phase gates]{\label{fig:z} \includegraphics[width=2.5in]{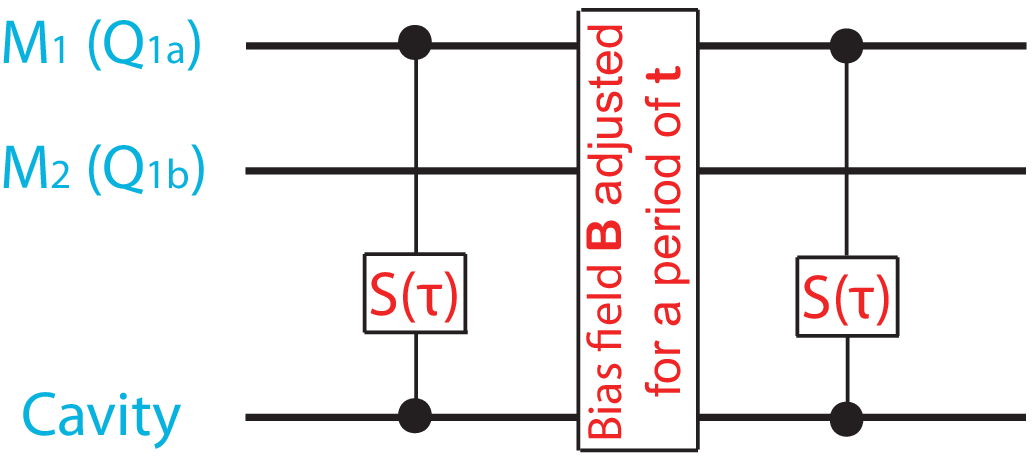}}
\end{array}$
\caption{Single qubit rotations in two independent directions, whose combinations constitute any single qubit operations for the dual-rail encoded qubits. Note that the cavity is assumed to be empty initially.}
\label{fig:singlequbitop}
\end{center}
\end{figure}

We encode the qubits in the dual-rail representation, where each logical qubit occupies two collective modes of the spin ensemble (see Fig.~\ref{fig:singlequbitop}). A logical qubit $\ket{Q_1}_L = \alpha \ket{0}_L + \beta \ket{1}_L$ corresponds to the physical state of modes $\ket{M_1 M_2}_M = \alpha \ket{10}_M + \beta \ket{01}_M$, where the ket notation denotes the number of excitations in the relevant mode for the physical qubits. Adopting a dual-rail encoding of course doubles the resource cost. However, the capacity of the ensemble quantum memory is of the order of $\sqrt{N}$; assuming $N = 10^{11}$, this is therefore not a practical constraint.  

An experiment would begin with cooling the spins so that there are a negligible number of excitations in the ensemble, which is then approximately in the ground state $\ket{{\bf 0}}$. For each logical qubit we then load the cavity with a single photon, and perform a {\it SWAP} gate between the cavity and the appropriate memory mode to represent $\ket{0}_L$. A logical qubit $\ket{Q}_L$ of arbitrary state can then be prepared through an appropriate single qubit rotation on the Bloch sphere.

Any single qubit rotation can be formed from a combination of rotations about two different axes~\cite{nielsen00}. In Fig.~\ref{fig:singlequbitop} we show how rotations about the x-axis, and the z-axis can be performed in the dual-rail enconding. Both these rotations involve {\it SWAP} operations between the memory modes and an initially empty cavity. 

\subsection{X Rotations}
Suppose that our logical qubit is initially in the general state $\ket{Q_1} = \cos \frac{\theta}{2} \ket{0}_L + e^{i \phi} \sin \frac{\theta}{2} \ket{1}_L$. We thus start with the initial state $\big(\cos \frac{\theta}{2} \ket{10}_M + e^{i \phi} \sin \frac{\theta}{2} \ket{01}_M \big)\ \ket{0}_C$, where C denotes the cavity state. After the first full {\it SWAP}, the physical state becomes 
\begin{equation}
\cos \frac{\theta}{2} \ket{00}_M \ket{1}_C + e^{i \phi} \sin \frac{\theta}{2} \ket{01}_M \ket{0}_C. 
\label{eq:swap}
\end{equation}
To implement a rotation around the x-axis on the Bloch sphere, a partial {\it SWAP} is now performed between the cavity and mode 2, resulting in the following state:
\begin{equation}
\hspace{-2.65cm} \cos \frac{\theta}{2}\ \big(\cos \theta' \ket{00}_M \ket{1}_C - i \sin \theta' \ket{01}_M \ket{0}_C \big) + e^{i \phi} \sin \frac{\theta}{2}\ \big( \cos \theta' \ket{01}_M \ket{0}_C - i \sin \theta' \ket{00}_M \ket{1}_C \big)
\end{equation}
or equivalently, 
\begin{equation}
a \ket{00}_M \ket{1}_C + b \ket{01}_M \ket{0}_C
\end{equation}
where 
\begin{equation*}
a = \cos \frac{\theta}{2} \cos \theta' - i e^{i \phi} \sin \frac{\theta}{2} \sin \theta'
\end{equation*}
\begin{equation*}
b=  - i \cos \frac{\theta}{2} \sin \theta' + e^{i \phi} \sin \frac{\theta}{2} \cos \theta'
\end{equation*}
and $\theta' = J t$. When the second full {\it SWAP} has completed, the empty cavity decouples from the qubit $\ket{Q_1} = a \ket{0}_L + b \ket{1}_L$, ignoring a non-detectable global phase. An {\it X} gate is implemented in this way when $t = \tau$, i.e., when $\theta' = \frac{\pi}{2}$.

\subsection{Z Rotations}
To obtain a phase gate, we begin in the same fashion and obtain the state in Eq.~\ref{eq:swap}. However, we now exploit the fact that applying a different magnetic bias field does not affect the phase evolution of the component $\cos \frac{\theta}{2} \ket{00}_M \ket{1}_C$ (which still acquires phase at the same rate $\epsilon$), since the cavity photon energy is not affected by the magnetic field. On the other hand, a change $\delta B$ in the bias field shifts the spin ensemble mode energy to $\epsilon + \delta (\delta B)$, which results in a phase evolution of $e^{- i (\epsilon + \delta) t}$ for the part $e^{i \phi} \sin \frac{\theta}{2} \ket{01}_M \ket{0}_C$ during the time $t$. Ignoring the undetectable global phase $e^{- i \epsilon t}$, after the second full {\it SWAP} the empty cavity again decouples from the qubit $\ket{Q_1} = \cos \frac{\theta}{2} \ket{0}_L + e^{i (\phi - \delta t)} \sin \frac{\theta}{2} \ket{1}_L$. The relative phase accumulated $e^{- i \delta t}$ manifests itself as a phase gate for the logical qubit, where the phase can be controlled by the time $t$ and the bias shift $\delta B$.

\section{The Measurement-assisted Entangling Scheme} \label{sec:entangle}
\begin{figure}[h]
\begin{center}
\includegraphics[width=5.5in]{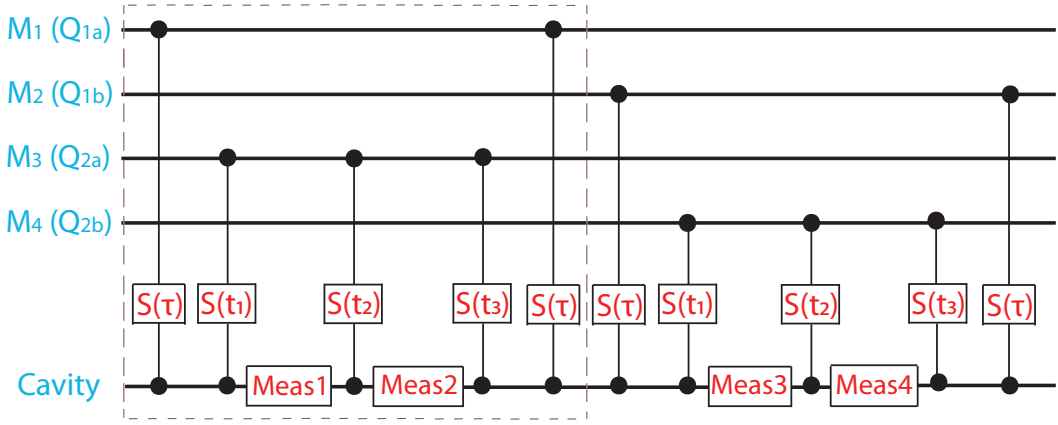}
\caption{The entangling operation for the dual-rail encoded qubits, implemented by a parity projection. A `building block' consists of all the operations inside the dashed box, and the parity projection requires two of such blocks. Each building block rules out the possibility of two excitations in the relevant two modes while preserving coherence between states with different occupation numbers. This requires two NO results for each block to the question whether there are 2 photons in the cavity or not. The initial cavity state is assumed to be empty, and the interaction times are chosen as follows: $t_1 = \frac{\pi}{4 J} \simeq 20$ ns, $t_2 = \frac{\pi}{2 J} \simeq 40$ ns and $t_3 = 2 \tau - t_1 - t_2 = \frac{\pi}{4 J} \simeq 20$ ns.}
\label{fig:entangle}
\end{center}
\end{figure}

Having established how to perform single qubit operations, we now require an entangling operation in order to produce a universal set of qubit gates~\cite{nielsen00}. In this section, we show how certain types of measurement can facilitate such an entangling scheme (see Fig.~\ref{fig:entangle}). In essence, we perform a measurement on the {\it parity} of two chosen logical qubits. If the parity is found to be odd \big( $\ket{01}_L$ or $\ket{10}_L$ \big), then we proceed with the protocol. Conversely, if the parity is found to be even \big( $\ket{00}_L$ or $\ket{11}_L$ \big), then we reject the state. This is therefore a probabilistic entangling operation, however, it is well established that such operations suffice for efficient universal quantum computing~\cite{benjamin09}. Our approach can be thought of as a filtering process; we detect the state $\ket{00}_L$ and reject it, and then similarly we detect and reject the state $\ket{11}_L$. Any state that passes through this process without rejection must be in the odd parity subspace \{$\ket{01}_L, \ket{10}_L$\}. Our protocol therefore consists of two basic blocks of operations, each ruling out the possibility of having two excitations in the relevant two modes while not destroying the quantum coherence between contributions to the state with different occupation numbers. As before, the cavity is emptied before the entangling gate operation, and it is tuned out of resonance with the ensemble modes during each measurement procedure.

Let us now consider the first block of operations, concerning modes $M_1$ and $M_3$ with the cavity. For each correctly encoded logical qubit, the two dual-rail modes together only contain a single excitation. However, for the $M_3$ mode-cavity interaction there is still the possibility of exchange of two excitations, originally from different qubits. This two-excitation exchange is governed by the Hamiltonian $H^{(2)}$ in Eq.~\ref{eq:2ex}, which has three distinct eigenvalues $2 \epsilon$, $2 (\epsilon - J)$ and $2 (\epsilon + J)$, with corresponding eigenvectors 
\begin{equation*}
\ket{v_0}_{MC} = \frac{1}{\sqrt{2}} \big(\ket{20} - \ket{02}\big)
\end{equation*}
\begin{equation*}
\ket{v_-}_{MC} = \frac{1}{2} \big( \ket{20} - \sqrt{2} \ket{11} + \ket{02}\big)
\end{equation*}
\begin{equation}
\ket{v_+}_{MC} = \frac{1}{2} \big(\ket{20} + \sqrt{2} \ket{11} + \ket{02}\big)
\end{equation}
respectively. Therefore, if the $M_3$ mode-cavity state after the first full {\it SWAP} is $\ket{11} = \frac{1}{\sqrt{2}} (\ket{v_+} - \ket{v_-})$, it evolves to $\frac{1}{\sqrt{2}} (\ket{v_+} + \ket{v_-}) = \frac{1}{\sqrt{2}} (\ket{20} + \ket{02})$ for an interaction time $t_1 = \frac{\pi}{4 J} \simeq 20$ ns. Performing a measurement designed to detect whether there are 2 photons in the cavity~\cite{johnson10} then removes the $\ket{02}$ component if the measurement result is a NO outcome. After a further exchange of $t_2 = \frac{\pi}{2 J} \simeq 40$ ns, the part $\ket{20}$ evolves to $\ket{02}$ which can be ruled out by a second measurement if the outcome is NO again. Thus, up to now when successful, we are assured that there will be at most one excitation between the modes $M_1$ and $M_3$, after it has been swapped back from the cavity. So far, we have only considered the two-excitation exchange, while there is also the exchange of one excitation. In order for the block to perform an identity operation, i.e., two {\it SWAP}s, on the one-excitation exchange subspace, a further interaction between the mode $M_3$ and cavity is required for a time of $t_3 = 2 \tau - t_1 - t_2 = \frac{\pi}{4 J} \simeq 20$ ns, followed by a full {\it SWAP} between the mode $M_1$ and cavity. 

Due to symmetry, the second block of operations has the same effect on modes $M_2$ and $M_4$, ensuring that there is also at most one excitation in between, if successful. Hence, starting with two correctly encoded qubits $\ket{Q_1 Q_2}_L = (\alpha_1 \ket{0}_L + \beta_1 \ket{1}_L) \bigotimes (\alpha_2 \ket{0}_L + \beta_2 \ket{1}_L)$ or physically $\ket{M_1M_2M_3M_4} = \alpha_1 \alpha_2 \ket{1010} + \alpha_1 \beta_2 \ket{1001} + \beta_1 \alpha_2 \ket{0110} + \beta_1 \beta_2 \ket{0101}$, successful implementation of the entangling gate leaves us with the mode state $\mathcal{N} (\alpha_1 \beta_2 \ket{1001} + \beta_1 \alpha_2 \ket{0110})$ or $\mathcal{N} (\alpha_1 \beta_2 \ket{0_{Q_1} 1_{Q_2}}_L + \beta_1 \alpha_2 \ket{1_{Q_1} 0_{Q_2}}_L) $ in the logical basis, where $\mathcal{N}$ is a normalisation constant. This is an odd-parity projection operation, which generates a maximally entangled {\it Bell} state when we start with $\ket{+}_{Q_1} \ket{+}_{Q_2}$ as the initial state of the logical qubits. Notably, such parity projections together with single qubit operations are adequate for universal QIP~\cite{benjamin09}.

\section{Errors}
With the above set of universal quantum gates, our integrated device is capable of performing probabilistic QIP within the electron spin ensemble, in the ideal case when no errors are present. We shall now consider the various error sources, namely, collective excitation decay, imprecise timing in cavity-mode swap, cavity photon leakage, and measurement errors.

As mentioned before, the decay rates of the collective spin excitations and that of cavity photon leakage are orders of magnitude smaller than the collective coupling $J$~\cite{wesenberg09}. Moreover, errors resulted from imperfect pulse timings are all linear, and timing precision with a nanosecond resolution is feasible~\cite{wesenberg09}. Therefore, these errors shall be treated negligible as compared to measurement errors~\cite{johnson10}, which shall be evaluated for the entangling gate in this section. We start with the logical qubit state
\begin{equation}
\ket{+}_{Q_1} \ket{+}_{Q_2} = \frac{1}{2} \big( \ket{1010} + \ket{1001} + \ket{0110} + \ket{0101} \big)_{M_1 M_2 M_3 M_4}
\label{eq:initial}
\end{equation}
with the cavity being $\ket{C} = \ket{0}$. There are two types of measurement errors to consider, corresponding to different incorrect outcomes to the question whether there are two photons in the cavity or not~\cite{johnson10}.

\subsection{Type I Measurement Error} \label{sec:1}
This type of error occurs when the measurement outcome gives a NO while the cavity is populated with two photons, and this occurs with an independent probability of $\eta_1$ for each measurement. It has an effect on the overall fidelity {\it F} of this entangling gate, reducing the degree of entanglement generated in the final state. With the cavity state $\ket{0}_C$ decoupled from the modes at the end of the protocol, the final density state for the modes $\ket{M_1M_2M_3M_4}$ in this case is (see Appendix)
\begin{equation}
\rho = \frac{1}{1+\eta_1} \bigg( \ket{\Psi^+}\bra{\Psi^+} + \frac{\eta_1}{2} (\sigma_1 + \sigma_2) \bigg)
\label{eq:fes}
\end{equation}
where
\begin{equation}
\ket{\Psi^+} = \frac{1}{\sqrt{2}} \bigg( \ket{1001} + \ket{0110} \bigg) \longrightarrow \frac{1}{\sqrt{2}} (\ket{01} + \ket{10})_{Q_1Q_2}
\label{eq:me}
\end{equation}
\begin{equation*}
\sigma_1 = \frac{1}{2} \bigg( \frac{(\ket{0020} - \ket{2000})(\bra{0020} - \bra{2000})}{2} + \ket{1010}\bra{1010} \bigg) 
\end{equation*}
\begin{equation}
\sigma_2 = \frac{1}{2} \bigg( \frac{(\ket{0002} - \ket{0200})(\bra{0002} - \bra{0200})}{2} + \ket{0101}\bra{0101} \bigg). 
\label{eq:sig}
\end{equation}

\begin{figure}[h]
\begin{center}$
\begin{array}{cc}
  \hspace{-3mm} \subfigure[]{\label{fig:fidel1} \includegraphics[width=3in]{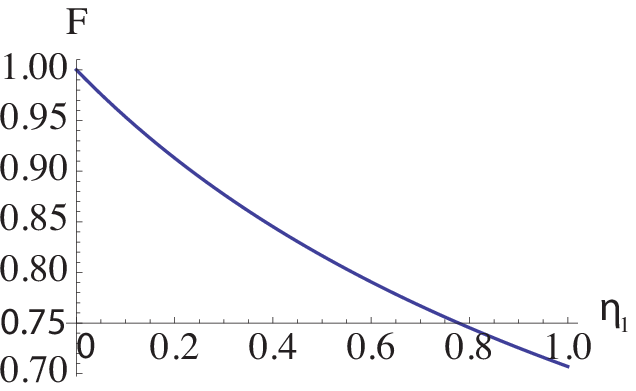}} & \hspace{5mm}
  \subfigure[]{\label{fig:eof1} \includegraphics[width=3in]{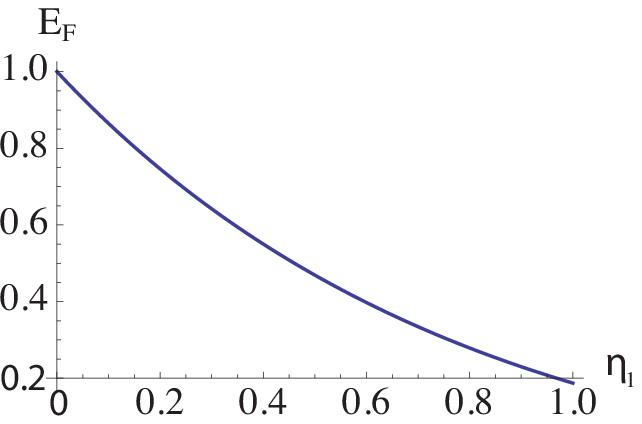}}
\end{array}$
\caption{Plots for the final state $\rho$ after the entangling operation, against the type I error $\eta_1$ for each measurement, showing (a) the fidelity $F$ with respect to $\ket{\Psi^+}\bra{\Psi^+}$; (b) the Entanglement of Formation $E_F$ when the population outside the computational subspace are simply ignored and projected out.}
\label{fig:fideof}
\end{center}
\end{figure}

In the ideal case where $\eta_1 = 0$, the resulting state after the entangling operation is the maximally entangled state $\ket{\Psi^+}$, as expected. For finite $\eta_1$ errors, the final mixed state $\rho$ describes that for bipartite quinits $Q_1$ and $Q_2$, where each quinit is a five-level system. Mapping the different physical states of the dual-rail modes $\ket{10}$, $\ket{01}$, $\ket{20}$, $\ket{02}$ and $\ket{00}$ to the logical quinit levels $\ket{0}$, $\ket{1}$, $\ket{2}$, $\ket{3}$ and $\ket{4}$, respectively for each $Q_i$, Eq.~\ref{eq:sig} becomes
\begin{equation*}
\sigma_1 \longrightarrow \frac{1}{2} \bigg( \frac{(\ket{42} - \ket{24})(\bra{42} - \bra{24})}{2} + \ket{00}\bra{00} \bigg)_{Q_1Q_2} 
\end{equation*}
\begin{equation}
\sigma_2 \longrightarrow \frac{1}{2} \bigg( \frac{(\ket{43} - \ket{34})(\bra{43} - \bra{34})}{2} + \ket{11}\bra{11} \bigg)_{Q_1Q_2} 
\label{eq:sigp}
\end{equation}
Although also being entangled in the levels $\ket{2}$, $\ket{3}$ and $\ket{4}$, we shall only be interested in the entanglement between the two levels $\ket{0}$ and $\ket{1}$, which span the computational basis space for conventional QIP tasks. Fig.~\ref{fig:fidel1} plots the fidelity $F$~\cite{jozsa94} of the resulting state $\rho$ (as given by Eq.~\ref{eq:fes}) with the ideal state $\ket{\Psi^+}\bra{\Psi^+}$ against the error $\eta_1$,
\begin{equation}
F (\rho, \ket{\Psi^+}\bra{\Psi^+}) = \Tr \bigg( \sqrt{\sqrt{\rho} \ket{\Psi^+} \bra{\Psi^+} \sqrt{\rho}} \bigg) = \sqrt{\frac{1}{1+\eta_1}}
\end{equation}
where $\ket{\Psi^+}$ is embedded in the five-level system by filling the levels that lie outside the computational basis with zero population. 

We see from Fig.~\ref{fig:fidel1} that a high fidelity is possible as long as the error $\eta_1$ remains small, however, this does not automatically guarantee the presence of entanglement for the bipartite quinits. We therefore apply an entanglement measure to see how the degree of entanglement changes as $\eta_1$ increases. The simplest (but questionable) way of doing so is to apply {\it Wootter}'s 
entanglement of formation $E_F$~\cite{wootters98} to the state $\rho$ projected onto the computational subspace, while ignoring the population in the other levels \big(see Fig.~\ref{fig:eof1}\big). The reason we consider this metric incorrect is that artificial entanglement is created by this projection process. Moreover, even small amounts of population in levels outside the computational subspace may result in a significant reduction in the degree of entanglement for the larger system in certain scenarios, as has recently been explored by Ref.~\cite{tiersch11}. 
Thus instead of the entanglement of formation, a more useful measure involving the whole five levels of the bipartite quinits is needed.

\begin{figure}[h]
\begin{center}$
\begin{array}{cc}
  \hspace{-2mm} \subfigure[]{\label{fig:comparison} \includegraphics[width=3in]{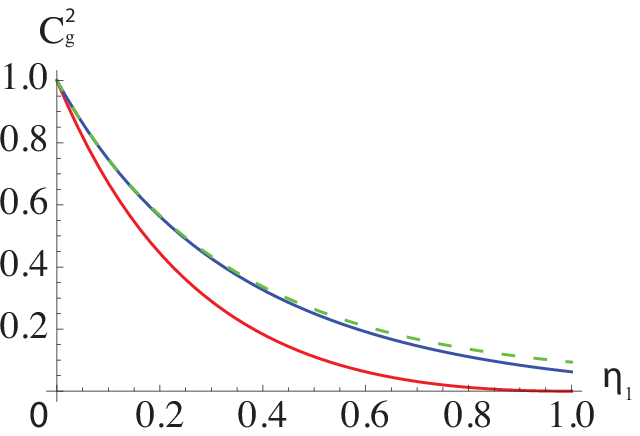}} & \hspace{3mm}
  \subfigure[]{\label{fig:eof} \includegraphics[width=3in]{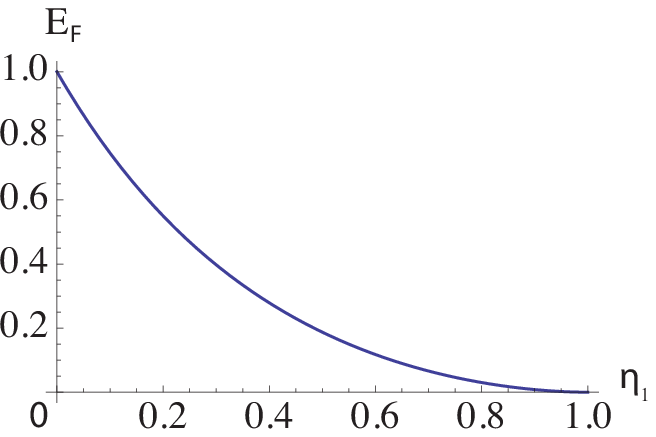}}
\end{array}$
\caption{(a) Plots of the generalised 2-concurrence $C_2^2$ in 5 dimensions against the $\eta_1$ error, for $\rho$ ({\it dashed green}), $\rho$ with dephased population outside the computational subspace ({\it solid blue}), and $\rho$ with all population discounted into the computational subspace ({\it solid red}); (b) Entanglement of formation $E_F$ plot for the discounted final state $\rho_d$ \big(red in (a)\big) against $\eta_1$.}
\label{fig:gce}
\end{center}
\end{figure}

We apply the generalised m-concurrence $C_m^2$ measure in $d$ dimensions as defined in~\cite{spengler10} and derived in~\cite{hiesmayr08}, where m denotes the number of parties involved and equals 2 in this case. In $d$ dimensions, $C_2^2$ ranges from zero for separable states to $\frac{2(d-1)}{d}$ for the maximally entangled generalised Bell states, such as $\frac{1}{\sqrt{d}} \sum_{i=0}^{d-1} \ket{ii}$. Moreover, it reduces to the Wootter's concurrence squared $C^2$~\cite{wootters98} when $d = 2$. Fig.~\ref{fig:comparison} illustrates the 2-concurrence ({\it dashed green}) of the final state $\rho$ (as in Eq.~\ref{eq:fes}) with $d=5$ against the error $\eta_1$. This is marginally lower than the projectively implemented Wootter's concurrence squared (not shown here), for each finite $\eta_1 \neq 0$. The difference is insignificant, partly due to the fact that the state $\rho$ in Eq.~\ref{eq:fes} is also entangled outside the computational space through the error components $\sigma_1$ and $\sigma_2$ (Eq.~\ref{eq:sigp}). Interestingly, even when $\eta_1 = 1$, the 2-concurrence is still greater than zero,  although no information is learnt at any of the four measurements. This can be understood as follows: we have assumed that the phase between the measured state and the rest of the superposition is lost through the interaction between the system and the measurement apparatus (see Appendix)~\cite{johnson10}. This corruption of phase information affects the subsequent evolution of the state and effectively causes a leakage of population from the components $\ket{00}_{Q_1Q_2}$ \big($\ket{11}_{Q_1Q_2}$\big) in the computational subspace to $\ket{42(3)}_{Q_1Q_2}$ and $\ket{2(3)4}_{Q_1Q_2}$. Entanglement then appears to exist -- and is registered by the 2-concurrence -- both within and outside the computational subspace. However, we stress that this apparent entanglement is useless for our purpose due to a complete lack of information about heralded failures, which prevents the required post-selection of successful runs of the protocol. 

As discussed above, our final state generally contains both useful and also `spurious'  entanglement. Therefore, we require a new metric for filtering out the latter kind to obtain a lower bound on the degree of useful entanglement. We shall achieve this by identifying a suitable theoretical operation that can be applied to the final state. It turns out to be insufficient to replace all population outside the computational basis with a completely mixed state in these levels, since the resulting curve \big({\it solid blue} in Fig.~\ref{fig:comparison}\big) still does not hit zero as $\eta_1 \rightarrow 1$. (In this case only the entanglement outside the computational subspace has been removed, but the entanglement inside of it is still present. Since no information has been learnt in the measurement, this entanglement is of the `spurious' kind which we want to discount.)
A better approach therefore proceeds as follows: we add any population that has leaked from the levels $\ket{00}_{Q_1Q_2}$ or $\ket{11}_{Q_1Q_2}$ back into computational subspace in a way that mixes the resulting state to the largest possible extent: 
\begin{equation}
\rho_d = \frac{1}{1+\eta_1} \bigg( \ket{\Psi^+} \bra{\Psi^+} + \frac{\eta_1}{2} \big( \ket{00} \bra{00} + \ket{11} \bra{11} \big) \bigg)_{Q_1Q_2}
\label{eq:rhod}
\end{equation}
The $C_2^2$ curve in 5 dimensions for this state then coincides with the square of Wootter's concurrence $C^2$ for two 2-dimensional systems, hitting zero as $\eta_1 \rightarrow 1$ \big({\it solid red} in Fig.~\ref{fig:comparison}\big). Indeed, this is what we would reasonably expect for maximal error $\eta_1$ when the measurement fails to report any useful information. 
Since only two levels are involved for $\rho_d$, we can also plot its entanglement of formation against $\eta_1$ in Fig.~\ref{fig:eof}, which is lower than that shown in Fig.~\ref{fig:eof1} as expected. We see that in order to achieve a high degree of useful entanglement as the result of the parity projection operation, $\eta_1$ has to be small. For example, $E_F (\rho_d) \simeq 0.75$ when $\eta_1 = 10$\% which is achievable with the current technology~\cite{johnson10}.

\subsection{Type II Measure Error} 
\begin{figure}[h]
\begin{center}$
\begin{array}{cc}
  \subfigure[$$ $\eta_1 = 0$ and 0.1]{\label{fig:ps1} \includegraphics[width=3in]{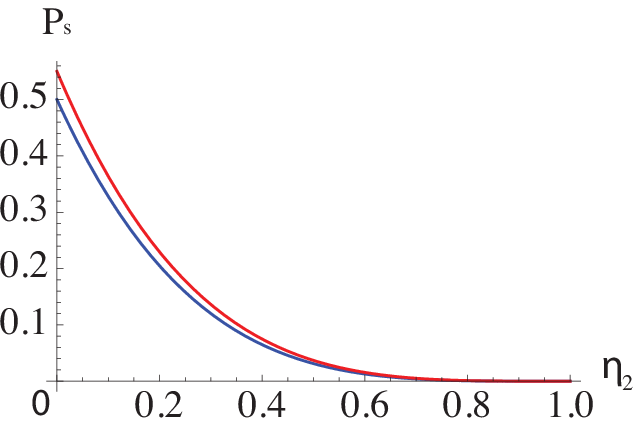}} & 
  \subfigure[$$ $\eta_2 = 0$]{\label{fig:ps2} \includegraphics[width=3in]{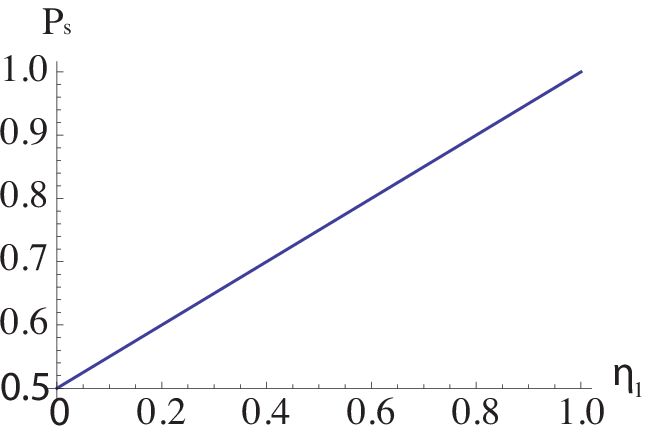}}
\end{array}$
\caption{The probability $P_f$ (a) for a type II error $\eta_2$ with $\eta_1 = 0$ (blue) and $\eta_1 = 0.1$ (red), respectively; (b) for a type I error $\eta_1$ with $\eta_2 = 0$. In the latter case, note that in spite of the increase in the probability $P_f$ the final state is still degraded as $\eta_1$ becomes larger.}
\label{fig:ps}
\end{center}
\end{figure}

This type of error occurs when a measurement outcome reports a YES while there are not two photons in the cavity, and this happens with an independent probability of $\eta_2$ for each measurement. For this kind of error, the entangling procedure is STOPPED and we have to start all over since a YES is reported. Consequently, this only affects the overall probability that we reach the end of the protocol without a heralded failure. We found this by considering the overall probability of passing all four measurement tests, and thus obtain
\begin{equation*}
\hspace{-2.5cm} P_f = \frac{7(1 - \eta_2) + \eta_1}{8} . \frac{(6+\eta_1)(1 - \eta_2) + \eta_1}{7+\eta_1} . \frac{(5+2\eta_1)(1 - \eta_2) + \eta_1}{6+2\eta_1} . \frac{(4+3\eta_1)(1 - \eta_2) + \eta_1}{5+3\eta_1} ~,
\end{equation*}
for the entire entangling operation (see Fig.~\ref{fig:ps}). 

In the absence of any error, the probability $P_f$ of reaching the end of the protocol is 50\%. $P_f$ decreases with $\eta_2$ for fixed $\eta_1$, and increases with $\eta_1$ for fixed $\eta_2$, although the quality of entanglement is of course reduced in this case. For the 10\% error in $\eta_1$ which gives rise to an 80\% concurrence, a type II error of $\eta_2 = 10$\%~\cite{johnson10} results in a probability $P_f \simeq 36.4$\%, i.e., roughly a third of the times.

\subsection{Cavity Photon Leakage}
Measurement errors are only important during the entangling scheme, while all other errors also affect the single qubit gate fidelities. Cavity photon loss is likely the next dominant error source, whereas timing errors and collective excitation decay can be considered negligible in comparison.

\begin{figure}[h]
\begin{center}
\includegraphics[width=4.5in]{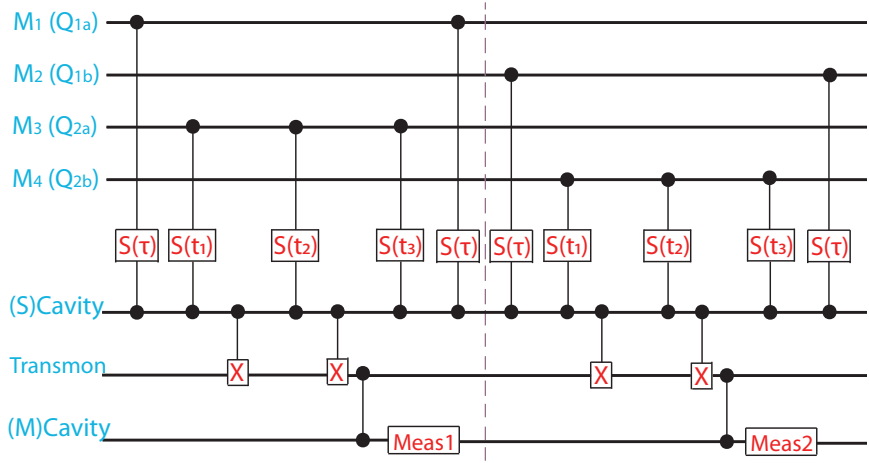}
\caption{The modified circuit for the whole integrated device during the entangling operation, where the number of measurements is halved. Nonetheless, the circuit still consists the same blocks of basic operations.}
\label{fig:entangle2}
\end{center}
\end{figure}

For this reason, we shall now briefly discuss an adaptation of our proposed circuit which can achieve a reduction of the cavity leakage effect. For each measurement procedure a Control-NOT ({\it C$_{NOT}$}) gate is first implemented on the transmon, which takes about 50 ns. The {\it C$_{NOT}$} gate is controlled by the number of photons in the cavity, so in this case the {\it C$_{NOT}$} gate is triggered whenever there are two photons in the cavity. The measurement cavity is then read out using a microwave probe, taking a time of about 400 ns~\cite{johnson10}. 

As the probe measurement thus takes a much longer time than the $C_{NOT}$ gates and the {\it SWAP} operations, we can adapt our earlier circuit to reduce the number of read-out steps required in each block of the entangling operation (see Fig.~\ref{fig:entangle2}). This modified block achieves the same effect as our previous circuit, i.e. it removes all components of the wavefunction with double excitations in the relevant two modes while not disturbing the remaining components of the state. For a cavity with photon decay time of $\frac{20}{2\pi} \simeq 3.2 \mu$s~\cite{johnson10} we expect a photon leakage of about 15.6\%  during  a 0.5 $\mu$s operation, but this source of error can be further suppressed by increasing the {\it Q}-factor of the storage cavity in the integrated device.

\section{Readout Scheme}
Full quantum state tomography can be carried out on the logical qubits to determine the dual-rail mode states. This is achieved through a sequence of {\it SWAP} operations followed by measurements on the storage cavity. This can also be extended to consider the entangled bipartite quinits as in Eq.~\ref{eq:fes}, by measuring the number of photons in the cavity swapped from each mode: Three measurements are then needed for each mode to determine whether there are zero, one, or two photons in the cavity.

\begin{figure}[h]
\begin{center}
\includegraphics[width=4.5in]{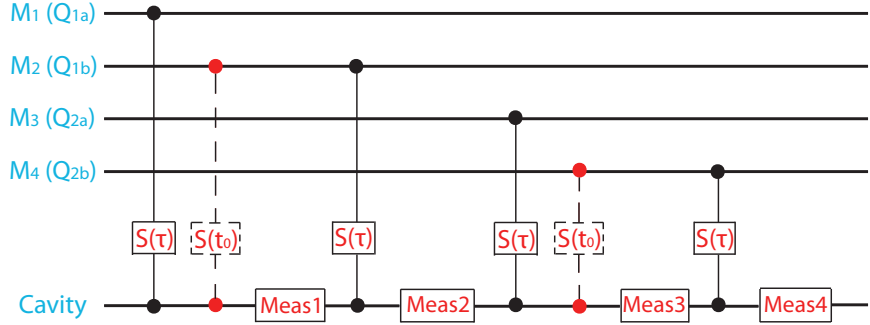}
\caption{Qubit measurement in two independent bases: (a) with only the solid ciruit, corresponding to measurement in the logical basis of \{$\ket{0}_{Q_i}$, $\ket{1}_{Q_i}$\}; (b) with the full circuit including the dashed gates, corresponding to measurement in the logical basis of \{$\ket{+}_{Q_i}$, $\ket{-}_{Q_i}$\} for $t_0 = \frac{\tau}{2}$.}
\label{fig:postsel}
\end{center}
\end{figure}

While the above-described full state tomography is straightforward to implement in principle, the number of measurements required for each run of the circuit might be challenging in a first experimental demonstration. Therefore, we propose the following alternative as a simpler verification of the entangling protocol: the aim is to use appropriate measurement statistics on the final state to reconstruct a density matrix that is consistent with the one given by Eq.~\ref{eq:rhod}. To achieve this, we start with repeated runs of the solid circuit in Fig.~\ref{fig:postsel}(a), with each measurement only checking whether there is a single photon in the cavity or not. An answer of NO means that there must have been either zero or two photons in the relevant mode $M_i$. Our default assumption in this case will be that the mode was in the state $\ket{0}_{M_i}$. Only when we obtain the total combination $\ket{0000}_{M_1M_2M_3M_4}$, do we assume that there must have existed two excitations in one of the four modes. However, such states only occur due to errors in the protocol; they lie outside the computational basis and do not contribute towards the final entangled resource state. As discussed in Sec.~\ref{sec:1} we account for these errors by adding their population to the computational subspace in a way that ensures that  the quality of the desired resource is not artificially boosted. Performing many runs of this circuit then allows us to fill the diagonal entries of the density matrix (in the \{$\ket{0}_{Q_i}$, $\ket{1}_{Q_i}$\} basis), and to compare them with Eq.~\ref{eq:rhod}.  

However, the above procedure alone is not sufficient to distinguish quantum entanglement from classical correlations. We therefore also carry out repeated runs of the complete circuit including the dashed gates in Fig.~\ref{fig:postsel}(b), corresponding to measurement in the logical basis of \{$\ket{+}_{Q_i}$, $\ket{-}_{Q_i}$\} for $t_0 = \frac{\tau}{2}$. This enables us to obtain the coherences of the density matrix, which can then once more be compared with Eq.~\ref{eq:rhod}. Note that in measuring the coherences, population outside the computational basis is no longer relevant, and can be directly discarded. Combining the two similar procedures described above, one can verify whether the entangled resource at the end of the protocol is consistent with the discounted qubit state $\rho_d$ for a known fixed $\eta_1$ error, up to a certain confidence level.

\section{Conclusion}
We have devised measurement-based entanglement protocol  for dual-rail encoded qubits in the collective excitations of an electron spin ensemble, coupled to a superconducting resonator linked with a measurement apparatus. An detailed analysis of the predominant error sources of the integrated device indicates that probabilistic QIP with the spin ensemble should be feasible with the current technology.

\section*{Acknowledgments}
We gratefully acknowledge helpful discussions with J. H. Wesenberg and H. Cable. This work was supported by the National Research Foundation and Ministry of Education, Singapore, and Hertford college, Oxford. We also acknowledge support from EPSRC (EP/F028806/1).

\section*{Appendix}
In this Appendix, we go through the evolution of the total mode-cavity state under the entangling operation with type I measurement error $\eta_1$, to derive the final state $\rho$ as in Eq.~\ref{eq:fes}. Since there are two excitations in the whole system, the phase evolution $e^{-2 i \epsilon t}$ is global and ignored at all times.

The initial total state $\ket{M_1 M_2 M_3 M_4 C}$ of the modes (Eq.~\ref{eq:initial}) and the cavity evolves to
\begin{equation}
\hspace{-1cm} \frac{1}{2} \bigg(- \frac{\ket{00200} + \ket{00002}}{\sqrt{2}} - i \frac{\ket{00011} - i \ket{00110}}{\sqrt{2}} + \frac{\ket{01100} - i \ket{01001}}{\sqrt{2}} + \ket{01010} \bigg)
\end{equation}
after the first full $M_1$C {\it SWAP} followed by the $M_3$C coupling $S(t_1)$. When the first measurement gives an outcome of NO, we are left with the density state
\begin{equation}
\hspace{-1cm} \rho_1 = \frac{8}{7+\eta_1} \bigg( \frac{7}{8} \ket{A_1}\bra{A_1} + \frac{1}{8} \eta_1 \ket{00002} \bra{00002} \bigg)
\label{eq:rho1}
\end{equation}
where 
\begin{equation}
\hspace{-1cm} \ket{A_1} = \sqrt{\frac{2}{7}} \bigg( \frac{-1}{\sqrt{2}} \ket{00200} + \frac{ -i \ket{00011} - \ket{00110}}{\sqrt{2}} + \frac{\ket{01100} - i \ket{01001}}{\sqrt{2}} + \ket{01010} \bigg)
\end{equation}
all normalised. Note that here with the set-up in~\cite{johnson10}, we have assumed that the measurement apparatus interacts with the cavity-ensemble system regardless of its ability in reporting correct answers. In so doing, even when $\eta_1 = 1$ where no information is to be learned from the measurement result, the state has still changed, namely, dephased according to Eq.~\ref{eq:rho1}.

The $M_3$C coupling $S(t_2)$ is a full {\it SWAP} for both the exchange of one and two excitations. Conditional on the second measurement giving an answer NO as well, the density state becomes 
\begin{equation}
\hspace{-1cm} \rho_2 = \frac{1}{2 (3 + \eta_1)} \bigg( 6 \ket{A_2} \bra{A_2} + \eta_1 \ket{00002} \bra{00002} + \eta_1 \ket{00200} \bra{00200} \bigg)
\label{eq:rho2}
\end{equation}
where
\begin{equation}
\hspace{-1cm} \ket{A_2} = \frac{1}{\sqrt{6}} \bigg( (- \ket{00110} + i \ket{00011}) + (-i \ket{01001} - \ket{01100}) + \sqrt{2} \ket{01010} \bigg)
\end{equation}
all normalised. 

The remaining operations in the first block results in the density state 
\begin{equation}
\hspace{-1cm} \frac{1}{3+\eta_1} \bigg( 3\ket{A_{f_1}} \bra{A_{f_1}} + \eta_1 \rho_0 \bigg)
\label{eq:rhof1}
\end{equation}
where
\begin{equation}
\hspace{-1cm} \ket{A_{f_1}} = \frac{1}{\sqrt{3}} \bigg( \ket{10010} - \ket{01100} + \ket{01010} \bigg),
\end{equation}
and
\begin{equation}
\hspace{-1cm} \rho_0 = \frac{1}{2} \bigg( \frac{(\ket{00200} - \ket{20000})(\bra{00200} - \bra{20000})}{2} + \ket{10100}\bra{10100} \bigg) 
\end{equation}
which does not evolve further during the rest of this entangling gate operation since the second block of operations never access the modes $M_1$ and $M_3$. At this point, the cavity is empty and decoupled from the modes. 

Similarly due to symmetry, upon the completion of the second block of operations on the modes $M_2$ and $M_4$, the cavity state $\ket{0}_C$ is again decoupled from the modes and we have for the modes $M_1M_2M_3M_4$ the density state (Eq.~\ref{eq:fes}-\ref{eq:sig})
\begin{equation*}
\hspace{-1cm} \rho = \frac{1}{1+\eta_1} \bigg( \ket{\Psi^+}\bra{\Psi^+} + \frac{\eta_1}{2} (\sigma_1 + \sigma_2) \bigg)
\end{equation*}
where
\begin{equation*}
\hspace{-1cm} \ket{\Psi^+} = \frac{1}{\sqrt{2}} \bigg( - \ket{1001} - \ket{0110} \bigg)
\end{equation*}
\begin{equation*}
\hspace{-1cm} \sigma_1 = \frac{1}{2} \bigg( \frac{(\ket{0020} - \ket{2000})(\bra{0020} - \bra{2000})}{2} + \ket{1010}\bra{1010} \bigg) 
\end{equation*}
\begin{equation*}
\hspace{-1cm} \sigma_2 = \frac{1}{2} \bigg( \frac{(\ket{0002} - \ket{0200})(\bra{0002} - \bra{0200})}{2} + \ket{0101}\bra{0101} \bigg). 
\end{equation*}

\end{document}